\documentclass[aps,twocolumn]{revtex4}
\usepackage{newlfont}
\usepackage{amssymb}
\usepackage{amsfonts}
\usepackage{amsmath}
\usepackage{wasysym}
\usepackage{graphicx}
\usepackage{bm}

\usepackage{graphicx}
\usepackage{epsfig}
\usepackage{newlfont}
\usepackage{amssymb}
\usepackage{amsfonts}
\usepackage{amsmath}
\usepackage{graphicx}
\usepackage{bm}

\usepackage{amsthm}

\begin{document}

\title{All Multiparty Quantum States Can Be Made Monogamous}

\author{Salini K.\(^{1}\), R. Prabhu\(^{2}\), Aditi Sen(De)\(^{2}\), and Ujjwal Sen\(^{2}\)}
\affiliation{\(^{1}\)School of Physics, IISER TVM, Thiruvananthapuram, Kerala, India\\
\(^{2}\)Harish-Chandra Research Institute, Chhatnag Road, Jhunsi, Allahabad 211 019, India}

\begin{abstract}
Monogamy of quantum correlation measures puts  restrictions on the sharability of quantum correlations in multiparty quantum states. Multiparty quantum states can satisfy or violate monogamy relations with respect to given quantum correlations. We show that all multiparty quantum states can be made monogamous with respect to all  measures. More precisely, given any quantum correlation measure that is non-monogamic for a multiparty quantum state, it is always possible to find a monotonically increasing function of the measure that is monogamous for the same state. The statement holds for all quantum states, whether pure or mixed, in all finite dimensions and for an arbitrary number of parties. The monotonically increasing function of the quantum correlation measure satisfies all the properties that is expected for quantum correlations to follow. We illustrate the concepts by considering a thermodynamic measure of quantum correlation, called the quantum work deficit.
\end{abstract}

\maketitle

\section{Introduction}
\label{sec:introduction}

Sharing  of quantum correlations among many parties is known to play an important role 
in quantum phenomena, ranging from quantum communication protocols \cite{BW, teleportation, exp, comm-review} to 
cooperative events in quantum many-body systems \cite{amader-AdP, Andreas-Fazio-Vlatko-RMP}.
It  is therefore important to conceptualize and quantify quantum correlations, for which investigations are usually pursued 
in two directions, viz. the 
entanglement-separability \cite{HHHH-RMP}  and the information-theoretic \cite{Modi} ones. Any such measure  of quantum correlation is expected to satisfy 
a monotonicity (precisely, non-increasing) under an intuitively satisfactory set of local quantum operations. 

For a quantum state which is shared between more than two parties, one may expect that all the
measures of quantum correlation  would additionally follow a  monogamy property \cite{Wootters, Bennetteof, KW, monogamyN},
which restricts the sharability of quantum correlations among many parties.
In the case of three parties, say, Alice, Bob and Charu, monogamy of a measure, ${\cal Q}$, says that the sum, ${\cal Q}_{AB}+{\cal Q}_{AC}$, of quantum correlations of the two-party local states between the Alice-Bob and the Alice-Charu pairs, should not exceed the quantum correlation, ${\cal Q}_{A:BC}$, of Alice with Bob and Charu taken together. 
Alice is therefore alloted a special status, and is called
the ``nodal observer''. 
If the tripartite state, shared between the three parties, Alice, Bob, and Charu, is symmetric under exchange of particles, then any of the three parties in the monogamy relation can act as the nodal observer. However, if the state under consideration is non-symmetric under interchange of particles, then we allot the status of the nodal observer to the party $i$ that minimises the monogamy expression ${\cal Q}_{i:jk}-{\cal Q}_{ij}-{\cal Q}_{ik}$, with $i,\, j,\, k$ being chosen from Alice, Bob, and Charu, and with no two of $i,\, j,\, k$ being equal. Let us mention however that our results hold with other choices of the nodal observer also.
The concept of monogamy has also been carried over 
to more than two extra-nodal observers.
Classical correlations certainly do not satisfy  a monogamy constraint \cite{monogamyN}.
The monogamous nature of quantum correlations plays a key role in the security of quantum cryptography \cite{cryptoRMP}. Moreover, monogamy of
quantum correlations has recently been used to study frustrated spin systems \cite{koteswar}. Surprisingly however, there are important and useful entanglement measures that do not satisfy monogamy for certain multiparty quantum states, an example being the entanglement of formation \cite{Bennetteof}, which quantifies the amount of entanglement required for preparation of a given bipartite quantum  state.
Nevertheless, it was found that for multiqubit systems, 
the concurrence squared \cite{concurrence},  a monotonically increasing function of the entanglement of formation 
is monogamous \cite{Wootters, Bennetteof, KW, monogamyN}. Similarly, the square of concurrence and entanglement of formation are monogamous for arbitrary multiqubit systems \cite{OliveiraBai}, although concurrence and entanglement of formation themselves are not so. Recently, it was shown that the information-theoretic quantum correlation measure, quantum discord \cite{discord1, discord2}, 
can violate monogamy \cite{amaderPrabhu, Giorgi, lightcone, RenFan} (cf. \cite{LOCC-monogamy, Dagmar}), and again a monotonically increasing function of the quantum discord satisfies monogamy for three-qubit pure states
\cite{monogamyDnew}. 

In this paper, we show that if any bipartite quantum correlation measure,
 of an arbitrary number of parties in arbitrary finite dimensions, is 
 non-increasing
 under loss of a part of a local subsystem, any multiparty quantum state is either already monogamous with respect to that measure 
 or an increasing function of the bipartite measure can make it so. 
Note that the result holds for both pure and mixed states.
It is interesting to note that the increasing function also satisfies all the properties for being a measure of quantum correlation,
which include monotonicity under local operations and vanishing for ``classically correlated'' states (which is the set of separable states for measures of entanglement). Moreover we show that the function can always be chosen to be reversible, so that there is no loss of information in applying the function on the parent quantum correlation \cite{HHHH-RMP, Vidal}. To illustrate the result, we show that although the quantum work-deficit \cite{workdeficit}, an information-theoretic quantum correlation measure, violates monogamy even for three-qubit pure states, the states become monogamous when one considers integer powers of the measure. In stark contrast to what happens for concurrence and quantum discord, we show that for the three-qubit generalized W states \cite{Wstate, dur-vidal-cirac}, the fourth power of quantum work-deficit is required to obtain monogamy for these states. In case of arbitrary three-qubit W-class states \cite{Wstate, dur-vidal-cirac} and the GHZ-class states \cite{GHZ, dur-vidal-cirac}, to obtain monogamy of quantum work-deficit, one requires higher polynomials. We also find that three-qubit pure states that are monogamous with respect to quantum discord are also so with respect to quantum work-deficit. 



\section{Turning non-monogamous multisite quantum states into monogamous ones}



Let ${\cal Q}$ be a quantum correlation measure that is defined for arbitrary bipartite states (pure or mixed) in arbitrary finite dimensions. 
Consider a three-party quantum state (pure or mixed), \(\varrho_{ABC}\), in arbitrary finite dimensions, shared between three observers, Alice \((A)\), Bob $(B)$, and Charu $(C)$.
Let \(\mathcal{Q}_{AB}\) denote the quantum correlation \(\mathcal{Q}\) for the two-party reduced state \(\varrho_{AB} = \mbox{tr}_{C}\varrho_{ABC}\). \(\mathcal{Q}_{AC}\) is similarly defined. 
Let \(\mathcal{Q}_{A:BC}\) denote the quantum correlation for the state \(\varrho_{ABC}\) in the \(A:BC\) partition. To prove our results, we consider quantum states of three finite dimensional systems. However, the results can be generalized to an arbitrary number of finite dimensional systems.

The measure \(\mathcal{Q}\) is said to satisfy monogamy for the state \(\varrho_{ABC}\) if 
\({\cal Q}_{A:BC} \geq {\cal Q}_{AB}+{\cal Q}_{AC}\). 
The idea is that a measure will be called monogamous for a certain shared quantum state if the amount of quantum correlations that Alice has with Bob and Charu separately would be smaller than what she has with her partners taken together. 
The measure will be called strictly monogamous for \(\varrho_{ABC}\) if 
\({\cal Q}_{A:BC} > {\cal Q}_{AB}+{\cal Q}_{AC}\). 
On the other hand, 
\({\cal Q}_{A:BC} < {\cal Q}_{AB}+{\cal Q}_{AC}\), 
will imply that the measure is non-monogamous for the corresponding state. 

It is interesting to note that the ``monogamy score"  ${\cal Q}_{A:BC}-{\cal Q}_{A:B}-{\cal Q}_{A:C}$ \cite{Wootters,VanishingDisc} can be used to quantify sharability of quantum correlations in tripartite quantum systems. Such quantities has been employed to detect regime changes in frustrated quantum many-body systems in experimental nuclear magnetic resonance substances \cite{koteswar}.

The following theorem demonstrates that the non-monogamous nature of any measure for any state can be transformed to a monogamous one (in fact, strictly so), by considering an increasing function of the measure.
Let \(\mathcal{R}\) be the set of all real numbers.\\
\noindent \textbf{Theorem 1:} If \({\cal Q}\) violates monogamy for an arbitrary three-party quantum state \(\varrho_{ABC}\) in arbitrary finite dimensions, 
there always exists an increasing function \(f:\mathcal{R}\to \mathcal{R}\) such that 
\begin{equation}
 f({\cal Q}_{A:BC}) > f({\cal Q}_{AB}) + f({\cal Q}_{AC}),
\end{equation}
provided that 
\({\cal Q}\) is monotonically decreasing under discarding systems and invariance under discarding systems occurs only for monogamy-satisfying states.
\\

\noindent \texttt{Proof:}  
Let us first  rename 
\[
{\cal Q}_{A:BC} = x, \quad {\cal Q}_{AB} = y, \quad {\cal Q}_{AC} = z,
\]
for notational simplicity.
Then the constraints in the premise of the theorem (non-monogamy and monotonicity of \({\cal Q}\)) can be rewritten as
\[ x < y+z, \quad    x > y > 0, \quad        x > z>0.\]
Hence it follows that 
$ 0 < \frac{y}{x} < 1$ and $   0 < \frac{z}{x} < 1 $\\
This implies that
\begin{equation}
\lim_{n \to \infty}\left( \frac{y}{x}\right)^n = 0, \quad \lim_{n \to \infty}\left( \frac{z}{x}\right)^n = 0 
\end{equation}
Hence $\forall\)  \(\epsilon > 0$, there exists positive integers $n_1( \epsilon), n_2( \epsilon)$ such that
 \begin{eqnarray}
 \left(\frac{y}{x}\right)^m  < \epsilon  \quad \forall  \mbox{ positive integers } m \ge n_1(\epsilon), \nonumber \\
\left(\frac{z}{x}\right)^m  < \epsilon \quad  \forall \mbox{ positive integers }  m \ge n_2(\epsilon).
 \end{eqnarray}
Let us now choose $\epsilon = \epsilon_1 < \frac{1}{2}$.
Therefore,  $\left(\frac{y}{x}\right)^m  < \epsilon_1 $ and  $\left(\frac{z}{x}\right)^m  <\epsilon_1$,  $\forall $ positive integers $m \ge n(\epsilon_1)$, where $n(\epsilon_1) = \max \{ n_1(\epsilon_1),n_2(\epsilon_1)\}$
Adding the inequalities, we have $\left(\frac{y}{x}\right)^m  + \left(\frac{z}{x}\right)^m  <2\epsilon_1 <1$,  $\forall $ positive integers $m \ge n(\epsilon_1)$.
Hence the proof.
\hfill $\blacksquare$

The above theorem can be generalized to an arbitrary number of parties (say, $N$) 
by choosing $\epsilon = \epsilon_1 < \frac{1}{(N-1)}$, whereby an inequality $x \leq \sum^{N-1}_{i=1}y_i$ (with $x>y_i>0,\, i=1,2,\ldots, N-1)$ will lead us to $\sum^{N-1}_{i=1}\left(\frac{y_i}{x}\right)^m<1$ for a suitably chosen $m$.

Note here that invariance under discarding part of a subsystem implying monogamy, holds for many quantum correlation measures, 
including entanglement of formation and concurrence for three-qubit systems and quantum discord in arbitrary finite dimensional
three-party states.
%
%
Note also that any positive power of a measure vanishes for the same class of states for which the original measure vanishes, so that 
the set of states that is indicated to be ``classical'' by the original measure, is invariant after the transformation of 
the original measure into the new one. 
Let us also mention here that if
a measure is monotonically non-increasing for a certain class of local operations (possibly assisted by classical communication between 
the parties), a positive integer power of the measure also has the same property. 
Specifically, for a measure \(\mathcal{Q}\) and a multiparty state \(\rho\),
\(\mathcal{Q}(\rho) \geq \mathcal{Q}(\Lambda(\rho)) (\geq 0)\) implies that 
\((\mathcal{Q}(\rho))^\alpha \geq (\mathcal{Q}(\Lambda(\rho)))^\alpha\)
for any positive \(\alpha\), where \(\Lambda\) represents a map that can be 
implemented by local quantum operations and classical communication.
%
Note that while the cases of vanishing \(x,y,z\) have been ignored in the proof,
they can be handled easily.

There is no guarantee that a given power that is instrumental in rendering a quantum correlation monogamous for Alice as the nodal observer, will also work for Bob or Charu as the nodal observer. However, the lowest common multiple of the these powers, corresponding to the three nodal observers, does the job.

There do exist examples of situations where a non-strict monotonically increasing function turns a non-monogamous quantum correlation into a monogamous one. However, they do not preserve all information about the original quantum correlation. In other words, for such functions, knowing $f({\cal Q}(\rho_{AB}))$ will not necessarily imply a knowledge of ${\cal Q}(\rho_{AB})$. This can make the $f({\cal Q}(\rho))$ to be drastically less useful in comparison to ${\cal Q}(\rho) $. We therefore want to restrict ourselves to strictly monotonically increasing functions. More specifically, we consider only ``reversible functions", i.e., function $f$ such that $f(x)$ can be used to find $x$ for all arguments $x$.

We now show that the class of monogamous states is closed under the operation of taking positive integral powers of the corresponding 
measure.\\
\textbf{Theorem 2:} If a quantum correlation measure is monogamous for a three-party quantum state, any positive integer power of the measure is also monogamous for the same state.\\

\noindent \texttt{Proof:}
The premise implies that \(x\geq y+z\). Then for any positive integer \(m\), we have 
\begin{equation}
x^m \geq \left(y+z\right)^m = \sum_{k=0}^m{m \choose k} y^k z^{m-k},
\end{equation}
 which in turn is \(\geq y^m + z^m\), as \(y\), \(z\) are non-negative. 
Hence the proof. \hfill \(\blacksquare\)


\section{On monogamy of quantum work-deficit}

We will now consider the monogamy properties of the information-theoretic quantum correlation measure, called quantum work-deficit (WD) \cite{workdeficit}.
In particular, this will help to illustrate 
that positive powers of a measure can lead to monogamous nature for a state, when the measure itself is not so.


We begin by relating the monogamy properties of quantum discord, quantum work-deficit, and entanglement of formation. Consider an arbitrary pure three-party state \(|\psi\rangle_{ABC}\).
Let us denote the quantum discord for 
the state \(\sigma_{AB} = \mbox{tr}_C |\psi\rangle\langle\psi|\) by \(D_{AB}\), where the measurement is performed by the observer \(B\). 
\(D_{AC}\) is similarly defined, with the measurement being performed by the observer \(C\). 
The entanglements of formation of \(\sigma_{AB}\) and \(\sigma_{AC}\) are denoted by \(E^f_{AB}\) and \(E^f_{AC}\) respectively. 
Similar notations are used for the different varieties of 
 the  quantum work-deficits, 
\(\Delta\), \(\Delta^\leftarrow\), and \(\Delta^\rightarrow\). 
See the Appendix for the definitions of these measures. 
\\


\noindent \textbf{Proposition 1:} For an arbitrary three-party pure state,
\(D_{AB} + D_{AC} + H(\{p^B_i\}) + H(\{p^C_j\})= E^f_{AB} + E^f_{AC}  + H(\{p^B_i\}) + H(\{p^C_j\}) \geq \Delta_{AB}^\leftarrow + \Delta_{AC}^\leftarrow \geq \Delta_{AB} + \Delta_{AC}\),
where \(H(\{p^B_i\})\) is the entropy produced by the measurement in \(B\), and similarly for  \(H(\{p^C_j\})\). 
\\

\noindent \texttt{Proof:} It can be obtained from  Ref. \cite{KW} that for an arbitrary pure state \(|\psi\rangle_{ABC}\), 
\begin{equation}\label{baddi-jaRd}
E^f_{AB} - \sum_i p_i^C S(I \otimes M_i  \varrho_{AC}I \otimes  M_i^\dagger /p_i^C) = 0,
\end{equation} 
where \(\{M_i\}\) forms the optimal measurement by the observer \(C\) and
\(p_i^C\)  are the corresponding probabilities.
Here \(S(\cdot)\) denotes the von Neumann entropy of its argument. 
Therefore,
\[E^f_{AB} + H(\{p_i^C\}) - S\left(\sum_i I \otimes M_i  \varrho_{AC} I \otimes M_i^\dagger\right) = 0,\]
where \(H(\cdot)\) denotes the Shannon entropy of the probability distribution in its argument. Here we assume that projective measurements attain optimality, which is conjectured to be the case for
rank-2 states in Ref. \cite{seikhane-hobe-dekha}. Consequently, 
\(E^f_{AB} + H(\{p_i^C\}) \geq \Delta_{AB}^\leftarrow + S(\varrho_{AB}) \geq \Delta_{AB}^\leftarrow \geq \Delta_{AB}\).
Hence the result.
\hfill \(\blacksquare\)

Performing measurements on the first parties will lead to 
\(2E^f_{BC}  + H(\{p^A_i\}) + H(\{q^A_j\}) \geq \Delta_{AB}^\rightarrow + \Delta_{AC}^\rightarrow \geq \Delta_{AB} + \Delta_{AC}\), where \(H(\{p^A_i\})\) (\(H(\{q^A_j\})\)) is the entropy 
produced in the measurement at \(A\) on \(\sigma_{AB}\) (\(\sigma_{AC}\)). 
\\

%
\noindent \textbf{Theorem 3:} For an arbitrary pure three-party quantum state \(|\psi\rangle_{ABC}\),  
quantum discord is monogamous whenever the quantum work-deficit, \(\Delta^\leftarrow\), is so.
\\

\noindent \texttt{Proof:} From the definitions of quantum discord and WD, we obtain
\begin{equation}
D_{AB} = S_B + \Delta_{AB} - H(\{p_i^B\}),
\label{eq:discord11111}
\end{equation}
where
 \(S_B\) is the von Neumann entropy of \(\sigma_{B} = \mbox{tr}_{AC}|\psi\rangle\langle\psi|\). 
Since \(S_B - H(\{p_i^B\}) \leq 0\), 
\(
D_{AB} \leq \Delta_{AB}^\leftarrow
\).
For states for which WD is monogamous, we have
\begin{equation}
D_{AB} + D_{AC} \leq  \Delta_{AB}^\leftarrow + \Delta_{AC}^\leftarrow \leq \Delta_{A:BC}^\leftarrow = S_A = D_{A:BC}.
\label{eq:discord_monogamy}
\end{equation}
Here we assume that the minimum of work-deficit and quantum discord are attained by the same measurement.    
It is easy to see that the theorem holds even if the first parties perform the measurements.
\hfill \(\blacksquare\)

The converse of the theorem does not hold, and by numerically searching over $10^5$ randomly-chosen pure three-qubit states, uniformly with respect to the Haar measure, we find that there are 35.788\% of three-qubit pure states for which WD is non-monogamous while quantum discord is monogamous, 6.975\% of them for which WD and quantum discord are both non-monogamous, and 57.237\% of them for which WD and quantum discord are both monogamous.

\subsection{Monogamy of quantum work-deficit for three qubit  states}

\begin{figure}[h]%
\centering
\resizebox{\columnwidth}{!}{
\includegraphics{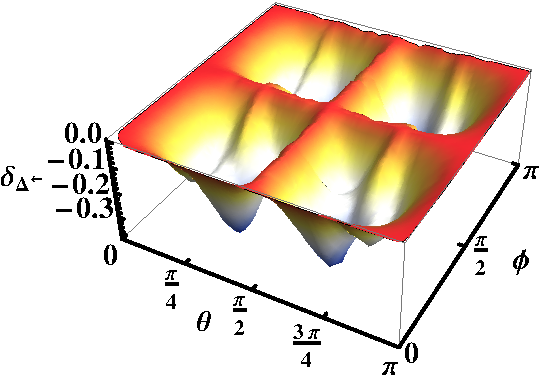}\hspace{3 em}
\includegraphics{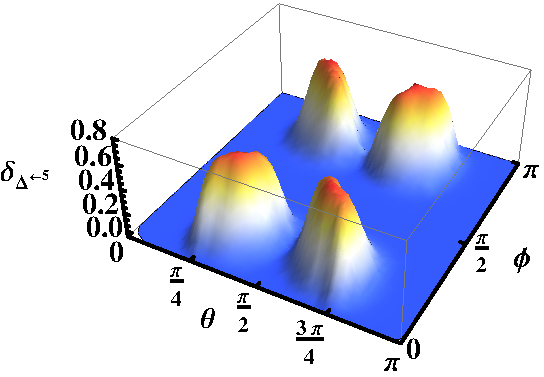}
}
\caption{Monogamy of quantum work-deficit. Left: The ``monogamy score'', \(\delta_{\Delta^\leftarrow}\), for quantum work-deficit is plotted, on the vertical axis, for the generalized W states, 
against the state parameters \(\theta\) and \(\phi\) on the base. 
Clearly, almost all generalized W states are non-monogamous with respect to quantum work-deficit as the quantum correlation measure. Right: All considerations are the same as in the left figure, except that 
the vertical axis represents the monogamy score \(\delta_{\left(\Delta^\leftarrow\right)^5}\) corresponding to the fifth power of WD.
As seen from the figure, almost all generalized W states are monogamous with respect to \(\left(\Delta^\leftarrow\right)^5\). 
(The vertical axes in both the figures are in qubits, while the base axes are dimensionless for both.)}%
\label{WDgenW}%
\end{figure}

We now consider the monogamy properties of 
quantum work-deficit for three-qubit pure states, and will begin by investigating the same for
an important class of three-qubit pure states, viz. the generalized W states \cite{Wstate, dur-vidal-cirac}, 
%
%
given by 
\begin{equation}
|\phi_{GW}\rangle=\sin \theta \cos \phi |011\rangle+\sin \theta \sin \phi |101\rangle+\cos \theta |110\rangle,
\label{eq:Wgen}
\end{equation}
where $\theta \in (0,\frac{\pi}{4}]$ and $\phi \in (0,2\pi]$.
Numerical evidence indicate that quantum work-deficit is non-monogamous for all or almost all members of this class (see Fig. \ref{WDgenW} (left)).
In other words, setting 
\begin{equation}
 \delta_\mathcal{Q} (\varrho_{ABC}) \equiv \mathcal{Q}_{A:BC} - \mathcal{Q}_{AB} - \mathcal{Q}_{AC} 
\end{equation}
for an arbitrary bipartite quantum  correlation measure \(\mathcal{Q}\) and an arbitrary three-party state \(\varrho_{ABC}\), we find that 
\begin{equation}
\delta_{\Delta^\leftarrow}(|\phi_{GW}\rangle) < 0
\end{equation}
for 
all the $10^4$ generalized W states that we randomly chose from the class of $|\phi_{GW}\rangle$. 
Note here that another information-theoretic quantum correlation measure, the quantum discord, 
can also be non-monogamous for these states \cite{amaderPrabhu, Giorgi, RenFan, lightcone}. However, recently it has been shown that the square of (one variety of) quantum discord is a monogamous quantity for all three-qubit pure states \cite{monogamyDnew}.
This however is no longer valid for WD. 
As stated in Theorem 1, suitably chosen integral powers of WD will be monogamous for any given state. And we find that for WD, monogamy for 
generalized W states is obtained (numerically)
for the fifth power (see Fig. \ref{WDgenW} (right)), i.e.   
\begin{equation}
 \delta_{\left(\Delta^\leftarrow\right)^5}(|\phi_{GW}\rangle) > 0
\end{equation}
for 
all the $10^4$ randomly chosen generalized W states. This feature remains unchanged when the measurement is performed by the observer $A$.

If one considers the W-class states, the percentage of non-monogamous states decreases slowly, as compared to the case of generalized W states with the increase of powers of work-deficit. In particular, we found by numerical simulation that the percentage of non-monogamous states with respect to  \({\left(\Delta^\leftarrow\right)^8}\)  is 10.76, decreasing from 100\% for  \(\Delta^\leftarrow\). The percentages are determined by Haar uniform generation of 10$^4$ randomly chosen states in the space of W-class states.

We have also considered the monogamy properties of general three-qubit pure states with respect to quantum work-deficit, \(\Delta^\leftarrow\). A histogram showing the relative frequencies of non-monogamous states among randomly chosen pure three-qubit states, for different powers of quantum work-deficit, is given in Fig. 2. Admixture of noise, if sufficiently small in amount, will still satisfy monogamy for the same power of $\Delta^\leftarrow$. Theorem 1 is however true for all mixed states, but larger levels of noise may require higher powers of $\Delta^\leftarrow$ to attain monogamy. In a given experimental setup, the experimenter can in principle find out her shared quantum state, and then Theorem 1 guarantees a finite positive power, $n$, for every bipartite quantum correlation measure, ${\cal Q}$, so that the corresponding ${\cal Q}^n$ will satisfy the monogamy relation.

\begin{figure}[h]%
\centering
\resizebox{\columnwidth}{!}{
\includegraphics{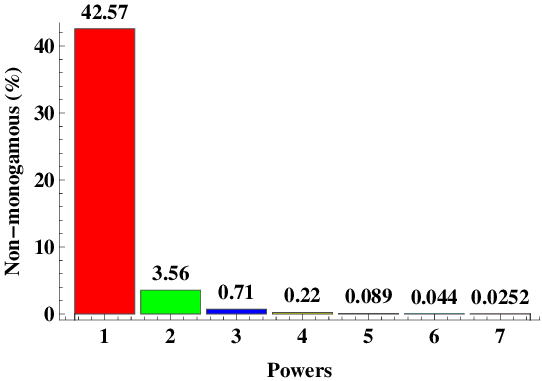}}
\caption{Relative frequencies of non-monogamous three-qubit pure states. 
We provide estimates of the percentages of the complete space of three-qubit pure states which violates monogamy with respect to quantum work-deficit and its integral powers. The histogram in the figure shows the percentages  on the vertical axis, while the different integral powers are on the horizontal axis. So, for example, the left-most (red) column indicates the estimated relative frequency of non-monogamous states with respect to the first power of WD, \(\Delta^\leftarrow\).
Both axes represent dimensionless parameters. The feature remains similar for the other variety of  WD, viz., \(\Delta^\rightarrow\), although in this case, the decrease of percentages is slower, with increasing powers of \(\Delta^\rightarrow\). The percentages are numerically determined by choosing $10^5$ three-qubit pure states Haar uniformly over the state space.}
\label{WDgenW1}%
\end{figure}


\section{Conclusion}

It is well-known that quantum correlation measures can be monogamous or non-monogamous for multisite quantum states. 
This can occur for quantum correlation measures of the entanglement-separability paradigm, as well as those of the information-theoretic one. We demonstrated that any quantum correlation measure that is non-monogamous for a multiparty quantum state can be made monogamous for the same by considering an increasing function of the measure. The transformed measure retains the important properties, like monotonicity under local operations and vanishing for ``classical'' states, of the original measure. We illustrate the results by using quantum work-deficit, an information-theoretic quantum correlation measure. We show that while the generalized W states are non-monogamous with respect to quantum work-deficit, the fifth power of the measure makes the states monogamous. We also discuss the monogamy properties of quantum work-deficit, and its powers, for arbitrary three-qubit pure states.

Let us mention here that in the literature, monotonically increasing functions of a quantum correlation measure are regarded with the same level of importance as the original measure. So, for example, 
the nearest-neighbor entanglement of quantum spin-1/2 systems \cite{amader-AdP, Andreas-Fazio-Vlatko-RMP} is usually investigated by employing the measure, concurrence, 
although a more physically meaningful measure is the entanglement of formation, with concurrence being an increasing function of the latter.

\section*{Acknowledgment}
RP acknowledges an INSPIRE-faculty position at the Harish-Chandra Research Institute (HRI) from the Department of Science and Technology, Government of India, and SK thanks HRI for hospitality and support.
\appendix

\section{Definitions of quantum correlation measures}

This appendix  provides a brief definition to the various quantum correlation measures used in this paper.

\section*{Entanglement of formation}

The entanglement of formation of a pure bipartite state, \(|\psi\rangle_{AB}\), shared between two parties \(A\) and \(B\),
 can be shown to be equal to the von Neumann entropy of 
the local density matrix of the shared state \cite{Bennetteof}: 
\begin{equation}
E(|\psi\rangle_{AB})= S(\varrho_A) = S(\varrho_B),
\end{equation} 
where $\varrho_{A}=\mbox{tr}_{B}|\psi\rangle\langle\psi|$ and similarly for \(\varrho_B\).
Entanglement of formation of a mixed bipartite state \(\rho_{AB}\) is 
then defined by the convex roof approach \cite{EoF1}: 
\begin{equation}
E(\rho)=\mbox{min}\sum_i p_iE(|\psi_i\rangle),
\end{equation}
where the minimization is over all pure state decompositions of $\rho = \sum_i p_i (|\psi_i\rangle \langle \psi_i|)_{AB}$.

\section*{Quantum discord}

Quantum discord is defined as the difference between two quantum information-theoretic quantities, whose classical counterparts are 
equivalent expressions for the classical mutual information
\cite{discord2,discord1}: 
\begin{equation}
Q(\rho_{AB})= {\cal I}(\rho_{AB}) - {\cal J}(\rho_{AB}).
\end{equation}
The ``total correlation'', \({\cal I}(\rho_{AB})\), of a bipartite state \(\rho_{AB}\) is given by \cite{qmi1} (see also \cite{Cerf1, GROIS1})
\begin{equation}
\mathcal{I}(\rho_{AB})= S(\rho_A)+ S(\rho_B)- S(\rho_{AB}),
\end{equation}
where $S(\varrho)= - \mbox{tr} (\varrho \log_2 \varrho)$ is the von Neumann entropy of the quantum state \(\varrho\), and 
 \(\rho_A\) and \(\rho_B\) are the reduced density matrices of  \(\rho_{AB}\).
On the other hand, \({\cal J}(\rho_{AB})\) can be interpreted as the amount of classical correlation in \(\rho_{AB}\), and is defined as 
\begin{equation}
 {\cal J}(\rho_{AB}) = S(\rho_A) - S(\rho_{A|B}). 
\end{equation}
Here
\begin{equation}
 S(\rho_{A|B}) = \min_{\{B_i\}} \sum_i p_i S(\rho_{A|i}),
\end{equation}
is the conditional entropy of \(\rho_{AB}\), conditioned on a measurement performed by \(B\) with a rank-one projection-valued measurement \(\{B_i\}\),
producing the states  
\(\rho_{A|i} = \frac{1}{p_i} \mbox{tr}_B[(\mathbb{I}_A \otimes B_i) \rho (\mathbb{I}_A \otimes B_i)]\), 
with probability \(p_i = \mbox{tr}_{AB}[(\mathbb{I}_A \otimes B_i) \rho (\mathbb{I}_A \otimes B_i)]\).
\(\mathbb{I}\) is the identity operator on the Hilbert space of \(A\).
\\

\section*{Quantum work-deficit}

We now briefly 
%
introduce the information-theoretic measure of quantum correlation, known as  quantum work-deficit \cite{workdeficit}  for an arbitrary bipartite quantum state $\rho_{AB}$. 
Let us begin by 
 considering the number, \(I_{G}\), of pure qubits that can be extracted from $\rho_{AB}$ by ``closed global operations'', with the latter consisting of any sequence of unitary operations and dephasing.
It can be shown that 
\begin{equation}
I_G (\rho_{AB})= N - S(\rho_{AB}),  
\end{equation}
where $N $ is the \(\log\) of  the dimension of the Hilbert space ${\cal H}$ on which $\rho_{AB}$ is defined.
This  thermodynamic ``work'' that can be extracted from the quantum state \(\rho_{AB}\) may require to employ global operations, which are not accessible to observers who are situated in separated laboratories. To obtain a
quantification of the amount of work that can be extracted from \(\rho_{AB}\) by local actions, 
we restrict   to ``closed local quantum operations and classical communication (CLOCC)'',
which consists of local unitaries, 
local dephasings, and sending dephased states from one party to another. Under these local actions,  the number of pure qubits that can be extracted 
is  given by
\begin{equation}
I_L(\rho_{AB}) = N - \inf_{\Lambda \in CLOCC} [S(\rho{'}_A) + S(\rho{'}_B)],
\end{equation}
where $S(\rho{'}_A) = \mbox{tr}_B (\Lambda (\rho_{AB}))$  and $S(\rho{'}_B) = \mbox{tr}_A (\Lambda (\rho_{AB}))$. 
For an arbitrary bipartite state $ \rho_{AB}$, the quantum work-deficit is then defined  as
\begin{equation}
\Delta(\rho_{AB}) = I_G(\rho_{AB}) - I_L(\rho_{AB}),
\end{equation}
and is interpreted as an information-theoretic quantum correlation measure of \(\rho_{AB}\). 
The quantity is not efficiently computable for arbitrary bipartite states. General CLOCC actions are also difficult to implement in an experiment. Therefore we will also consider the quantity \(\Delta_{AB}^\rightarrow\), in which 
we 
restrict our attention to CLOCC consisting of projection measurements at the single party (\(A\)) only for extracting work with local actions.
If the measurement is performed by \(B\), we denote it as \(\Delta_{AB}^\leftarrow\). 





\end{document}